\documentclass[11pt,a4paper]{article}
\usepackage{fullpage}
\usepackage {color}
\usepackage[pdftex]{graphicx}
\usepackage{amsmath}
\usepackage{amsfonts}
\usepackage{amssymb}
\usepackage{epstopdf}
\usepackage{color}

\begin{document}
\begin{flushright}
{ULB-TH/13-04}
\end{flushright}
\vskip 1cm
\begin{center}
{\huge Precision vs discovery: a simple benchmark}
\vskip .7cm
{\large C\'ecile Caillol${}^a$},
{\large Barbara Clerbaux${}^a$},
{\large Jean-Marie Fr\`ere${}^b$} and
{\large Simon Mollet${}^b$}
\vskip .7cm
\emph{
${}^a$ Interuniversity Institute for High Energies (IIHE),\\
Physique des particules \'el\'ementaires,\\
Universit\'e Libre de Bruxelles, ULB, 1050, Brussels, Belgium
\vskip 0.2cm
${}^b$ Service de Physique Th\'eorique,\\
Universit\'e Libre de Bruxelles, ULB, 1050 Brussels, Belgium}
\end{center}

\newcommand{\beq}{\begin{equation}}
\newcommand{\eeq}{\end{equation}}
\newcommand{\nn}{\nonumber}

\begin{abstract}

The discovery of the Standard Model Scalar Boson (Brout-Englert-Higgs particle) opens a new field of research, namely the structure of the scalar sector. Numerous extensions exist, and imply extra particles, including additional scalars. Two main alleys are open to investigate such deviations from the minimal standard model: precision measurements could indicate a deviation from the usual expectations, or a direct discovery of new scalar partners (or other particles) would establish an alternative. In this short note, we concentrate on a very simple model, for which  the respective reaches of the two approaches can be compared. This note also provides a strong incentive to pursue searches for extra "Standard Model Scalar-like" particles in the whole available energy spectrum.

\end{abstract}

\bigskip

\section{Introduction}

A new scalar has been found at LHC, with a mass around 125 GeV. To check that this is indeed the minimal version of the Brout-Englert-Higgs \cite{Brout:1964}\cite{Higgs:1964} particle in the Standard Model \cite{Weinberg:1967}, we can either rely on increasingly precise measurements comparing branching ratios and production rates, or satisfy ourselves that no further scalar structure is found. Both are, of course, open-ended tasks, as extensions  of the Standard Model scalar structure can in principle be arbitrarily close to the basic version.

To compare the approaches, we rely on a very simple model. Its initial version dates back more than 30 years \cite{Hill:1987}, but it is found as an ingredient in a number of more elaborate constructions, including the Next to Minimal Supersymmetric Standard Model (NMSSM) \cite{gavela}.

The principle of the model is very simple, and we will describe it quickly, before providing the corresponding Lagrangian: add a (real) singlet scalar $S$  boson to the Minimal Standard Model doublet $H$. As a singlet under the gauge group, it does not interact with the gauge bosons, and has no direct coupling to the fermions. Its only couplings to known particles are through the scalar. After symmetry breaking, the usual "Standard Model Scalar" (BEH boson) $H$ mixes with the neutral singlet. We thus end up with two mass states (2 "peaks"), which correspond to two neutral scalars:
\begin{eqnarray}
H_1 &=& H \cos\alpha + S \sin\alpha    \label{mixing1} \\
H_2 &=& - H \sin\alpha + S \cos\alpha . \label{mixing2}
\end{eqnarray}
$H_1$ has mass $m_1$ and $H_2$ has mass $m_2$, and we will by convention take $m_1=$ 125 GeV, while $m_2$ can be heavier or lighter. Since the singlet $S$ possesses no interaction of its own, the $H_1$ and $H_2$ states interact only by their $H$ component.

Note that in the balance between precision measurements and direct discovery, it should be kept in mind that the present example is a kind of "worst case" for the latter approach. Indeed, the particle added to the standard model exhibits no new interaction of its own, and only interacts through mixing. More general extensions would involve new interactions, which could boost production of the new components.

For simplicity, we will use "SMS" for the canonical Standard Model Scalar. In terms of production, the $H_1$ and $H_2$ particles are produced each exactly like a SMS of the corresponding mass, but with factors $\cos^2\alpha$ and $\sin^2\alpha$ respectively. For their decay, the branching ratios are identical to those of a hypothetical SMS of the same mass $m_1$ or $m_2$. Of course, the total width of the state, like the production rate, are weighted by $\cos^2\alpha$ and $\sin^2\alpha$ respectively, but this does not affect directly the observation rates in the various channels (once produced, unless $\alpha$ is vanishingly small, both particles decay quickly). One indirect effect might appear in the width of the peak, when this is observed -- this would facilitate the detection of heavy scalars, but probably requires the adaptation of the search.

It is thus fairly straightforward to exploit the existing searches for a SMS at various energies (and the corresponding bounds) to explore this model. For the time being, we will consider the parameters $m_1=$ 125 GeV, $m_2$ and $\alpha$ as independent (see more details below).

The situation is however complicated by the opening of the channel $H_2 \rightarrow 2 H_1$ for $m_2 \geq 2 m_1$ \cite{Basso:2013}. The branching ratio $BR(H_2\rightarrow 2H_1)$ can be determined in terms of the Lagrangian parameters, or in terms of masses and $\sin^2\alpha$.

We can thus present the existing data (possibly channel by channel, but preferably combinations, to achieve the best sensitivity in a large energy range) in different ways within the context of the model. In order to compare the "reach" of the precision measurement techniques, we plot the limits obtained on $\cos^2\alpha$ and $\sin^2\alpha$ as ordinates and abscissae respectively. The minimal model considered here is then represented by the line $\cos^2\alpha+\sin^2\alpha=1$, and it becomes easy to check which constraint (departure from expected rate at 125 GeV or limit on an extra peak at another energy) is the most restricting. Of course, upper limits on $\sin^2\alpha$ will depend on the energy range considered, and a series of exclusion lines will need to be drawn.

\begin{figure}[h]
\centering
\includegraphics[scale=0.55]{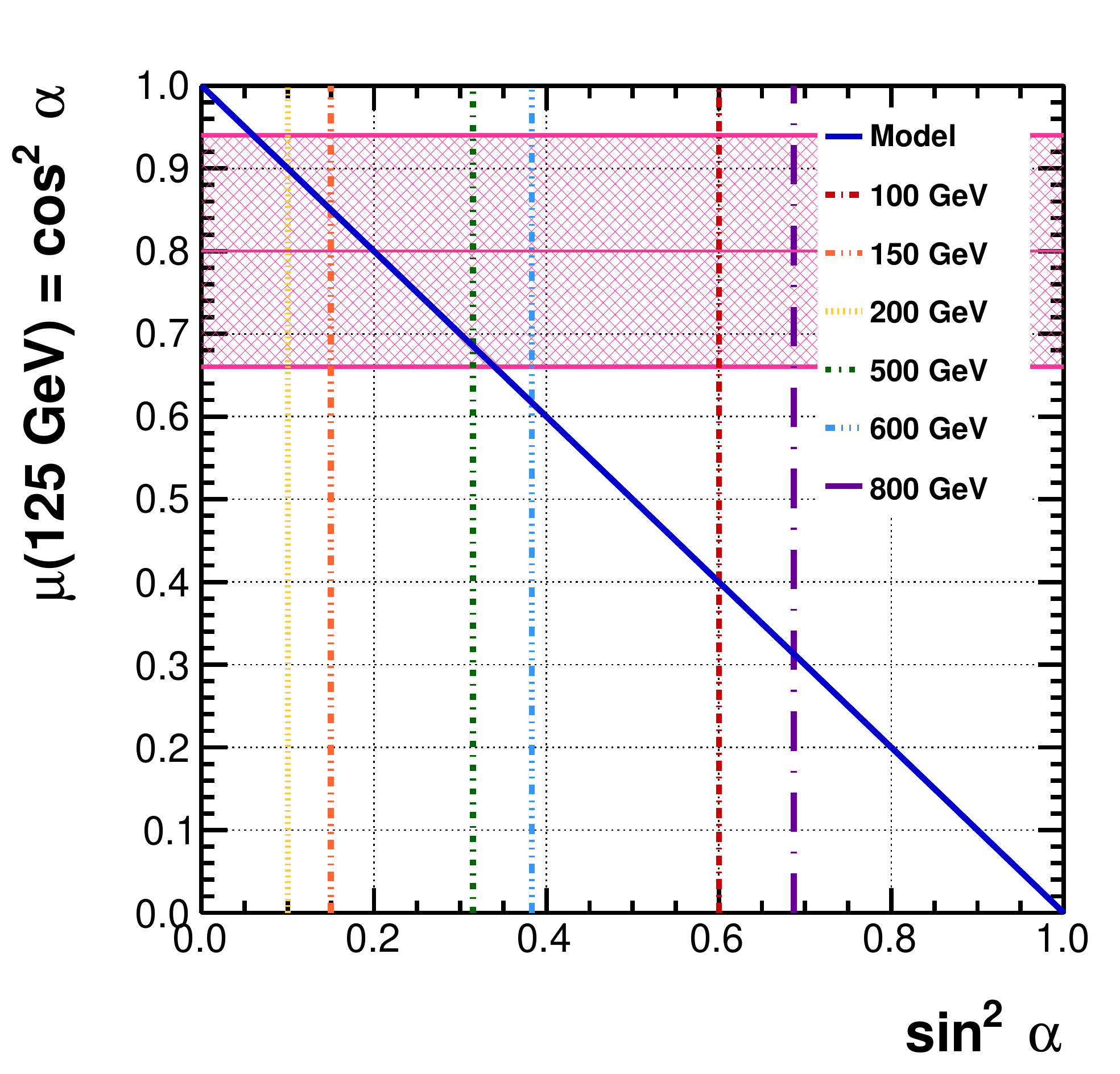}
\caption{The proposed benchmark compared to current public CMS data (HCP 2012) \cite{CMS:2012,CMS-PAS-HIG-13-005,CMS-PAS-HIG-13-014,CMS:5353,CMS:1129}. The extra singlet benchmark is constrained to be on the blue line $\sin^2\alpha + \cos^2\alpha=1$. The ordinates correspond to the precision measurement of the production of the 125 GeV SMS $\frac{(\sigma\times BR)_{\text{obs}}}{(\sigma\times BR)_{\text{exp}}}$, the pink hatched area represents the current $1\sigma$ confidence interval. The abscissa gives current production limits (this time at the more stringent 95$\%$ C.L.) for SM-like scalars of different masses (100 GeV, 150 GeV, 200 GeV, 500 GeV, 600 GeV and 800 GeV).\label{fig:benchmark}}
\end{figure}

We have performed the exercise in Fig.\ref{fig:benchmark}, using combination of currently available (public CMS data) \cite{CMS:2012,CMS-PAS-HIG-13-005,CMS-PAS-HIG-13-014,CMS:5353,CMS:1129}. Unfortunately the confidence levels of published direct searches (95$\%$ C.L.) and production rates (1$\sigma$) are different. We have plotted them as such (since we don't know the exact $\chi^2$), but one should keep in mind that the horizontal band corresponding to the production rate should be considerably broadened for a fair comparison to the direct searches. The result is nevertheless quite interesting, as it shows that, except for the lowest (below $\sim$ 110 GeV) and highest (above $\sim$ 600 GeV) values of the masses, the direct search is most constraining, despite being disfavoured by the specific model. We hope that this comparison can serve  to plot the expectations of future LHC runs, upgrades and alternative machines.

An alternate way to plot the constraints of this simplistic model consists in superimposing the upper bound on $\sin^2\alpha$ obtained from the 125 GeV scalar production rate as an horizontal line on the usual SM-like search plot $\frac{(\sigma\times BR)_{\text{obs}}}{(\sigma\times BR)_{\text{exp}}}$ (adapted to take into account the opening of the $H_2\rightarrow 2 H_1$ channel, see below) vs mass of the extra scalar; it provides a more synoptic view, but insists less on the complementarity of the approaches in narrowing the model parameters.

\section{Simple models}

We have treated this far the two main parameters (the mixing angle $\alpha$ and the second scalar mass $m_2$) as independent. We will show below that some dependence exists (as implied by decoupling requirements), but that it can be safely ignored at this stage. For this purpose, we now present an explicit form of the (scalar part of the) Lagrangian\footnote{A detailed study of the scalar potential stability in a related model can be found in \cite{Pruna:2013}.} -- to be further referred to as \textit{Extra Singlet Model (ESM)}. In \cite{Hill:1987}, the authors introduce the following Lagrangian ($\Phi$ and $\chi$ are the doublet and the singlet respectively)\footnote{The present normalization differs by an (arbitrary) factor of 2 from the one of reference \cite{Hill:1987}.}:
\beq
\mathcal{L}=\vert D_{\mu} \Phi\vert^2 +\frac{1}{2}\left(\partial_{\mu}\chi\right)^2
-\frac{\lambda_1}{2}\left(\vert\Phi\vert^2-f_1^2/2\right)^2
-\frac{\lambda_2}{2}\left(\vert\Phi\vert^2-f_2 \chi\right)^2.\label{Lag}
\eeq
It is an easy task to check that $\vert\Phi\vert^2=f_1^2/2$ and $\chi=f_1^2/2f_2$ minimize the potential. Then we can identify $f_1$ with $v\approx$ 250 GeV, the usual VEV of the SMS field, linked to the $W$ mass. The mass eigenvalues are then:
\beq
m_{\pm}^2=\frac{1}{2}\left(\lambda_2 f_2^2 + v^2 \lambda_3\right)\pm
\sqrt{v^2\lambda_2^2f_2^2+\frac{1}{4}\left(\lambda_2 f_2^2-v^2\lambda_3\right)^2},\label{masses}
\eeq
with $\lambda_3=\lambda_1+\lambda_2$. One can then show that the mixing angle is given by (both for $m_1>m_2$ and $m_2>m_1$):
\beq
\sin^2\alpha = \frac{\lambda_3-\left(m_1/v\right)^2}{\left(m_2/v\right)^2-\left(m_1/v\right)^2}.\label{sinSq}
\eeq
As expected, decoupling is achieved in the limit where $m_2$ becomes much larger than $m_1$ with bounded $\lambda_3$, as the mixing vanishes accordingly.

As mentioned before, the channel $H_2 \rightarrow 2 H_1$ opens for $m_2 \geq 2 m_1$. Its branching ratio can be evaluated in terms of masses and $\sin^2\alpha$ and its presence must be taken into account to constraint $\sin^2\alpha$.

With the new channel opens, the ratio $\frac{(\sigma\times BR)_{\text{obs}}}{(\sigma\times BR)_{\text{exp}}}$ becomes:
\beq
\frac{(\sigma\times BR)_{\text{obs}}}{(\sigma\times BR)_{\text{exp}}}=
\frac{s^2_{\alpha}\sigma_{\text{SM}}(m_2)}{\sigma_{\text{SM}}(m_2)}\times \frac{s^2_{\alpha}\Gamma_{\text{SM}}(m_2)}{s^2_{\alpha}\Gamma_{\text{SM}}(m_2)+\Gamma_{2H_1}(m_2)}=
\frac{s^4_{\alpha}\Gamma_{\text{SM}}(m_2)}{s^2_{\alpha}\Gamma_{\text{SM}}(m_2)+\Gamma_{2H_1}(m_2)}\equiv \mu(m_2).\label{corrMu}
\eeq
where $\Gamma_{2H_1}$ takes the new channel into account. When $m_2<2m_1$, $\Gamma_{2H_1}=0$ and we recover the simple result $\frac{(\sigma\times BR)_{\text{obs}}}{(\sigma\times BR)_{\text{exp}}}=s^2_{\alpha}$. Then the value of $s^2_{\alpha}$ is directly extracted from data in $\mu_{\text{data}}$. When the new channel is open, $\Gamma_{2H_1} \neq 0$ will modify this simple conclusion. At tree level $\Gamma_{2H_1}$ is given by:
\beq
\Gamma_{2H_1}=\frac{1}{16\pi m_2}\sqrt{1-\frac{4m_1^2}{m_2^2}}|\mathcal{M}(H_2\rightarrow 2 H_1)|^2.
\eeq
Obviously the matrix element $|\mathcal{M}|^2$ is model dependent and in the following we will restrict ourselves to the ESM.

To compute $\Gamma_{2H_1}$, we need to extract the $H_1^2 H_2$ from the Lagrangian. This comes from the cubic part of the potential:
\beq
V_{3}=\frac{\lambda_3}{2}v H^3 -\frac{\lambda_2}{2}f_2 H^2 S,
\eeq
and after rotation in the mass basis, the interesting contribution is:
\beq
V_{H_1^2H_2}=\left(-\frac{3}{2}\lambda_3 v c^2 s -\frac{1}{2}\lambda_2 f_2 c (1 - 3  s^2) \right) H_1^2H_2\equiv F H_1^2 H_2,
\eeq
which yields $|\mathcal{M}|^2=4F^2$.

Now let us replace $\lambda_3$, $\lambda_2$ and $f_2$ by their expressions in terms of masses and mixing. $\lambda_3$ has already been computed. For the two others, we have:
\beq
\lambda_2=\frac{s^2_{\alpha}c^2_{\alpha}(m_2^2-m_1^2)^2}{v^2(s^2_{\alpha}m_1^2+c^2_{\alpha}m_2^2)}
\eeq
and
\beq
f_2=v\frac{s^2_{\alpha}m_1^2+c^2_{\alpha}m_2^2}{s_{\alpha}c_{\alpha}(m_2^2-m_1^2)}.
\eeq
This gives\footnote{Our result differs from the one obtained in the current version of \cite{Basso:2013}, but the conclusions remain in  qualitative agreement.}:
\beq
F=-\frac{c^2 s}{2 v}(2m_1^2+m_2^2).
\eeq
Therefore $\Gamma_{2H_1}$ is given by:
\beq
\Gamma_{2H_1} =\frac{c^4_{\alpha}s^2_{\alpha}}{16\pi m_2 v^2}\sqrt{1-\frac{4m_1^2}{m_2^2}}(2m_1^2+m_2^2)^2 \equiv G(m_2) c^4_{\alpha}s^2_{\alpha},\label{dG}
\eeq
and is thus entirely determined in terms of masses and $\sin^2\alpha$. To extract the value of $s^2_{\alpha}$ from the data, we must solve (\ref{corrMu}) with $\Gamma_{2H_1}$ given by (\ref{dG}) and $\mu$ by $\mu_{\text{data}}$. If we define $S\equiv s^2_{\alpha}$, we have for any $m_2$:
\beq
\mu_{\text{data}}(m_2) G(m_2) S^2 - (\Gamma_{\text{SM}}(m_2)+2\mu_{\text{data}}(m_2) G(m_2)) S +\mu_{\text{data}}(m_2)(\Gamma_{\text{SM}}(m_2)+G(m_2))=0.
\eeq
Since $s^2_{\alpha}\leq 1$, we get:
\beq
s^2_{\alpha}=\gamma+1-\sqrt{\gamma^2+2\gamma(1-\mu_{\text{data}})},\label{newSin2}
\eeq
where $\gamma \equiv \frac{\Gamma_{\text{SM}}}{2\mu_{\text{data}} G}$. Note that $\mu<1$ by definition. When $\mu_{\text{data}}>1$, we cannot constraint the model. This is coherent with the relation (\ref{newSin2}).\\

\begin{figure}[h]
\centering
\includegraphics[scale=0.75] {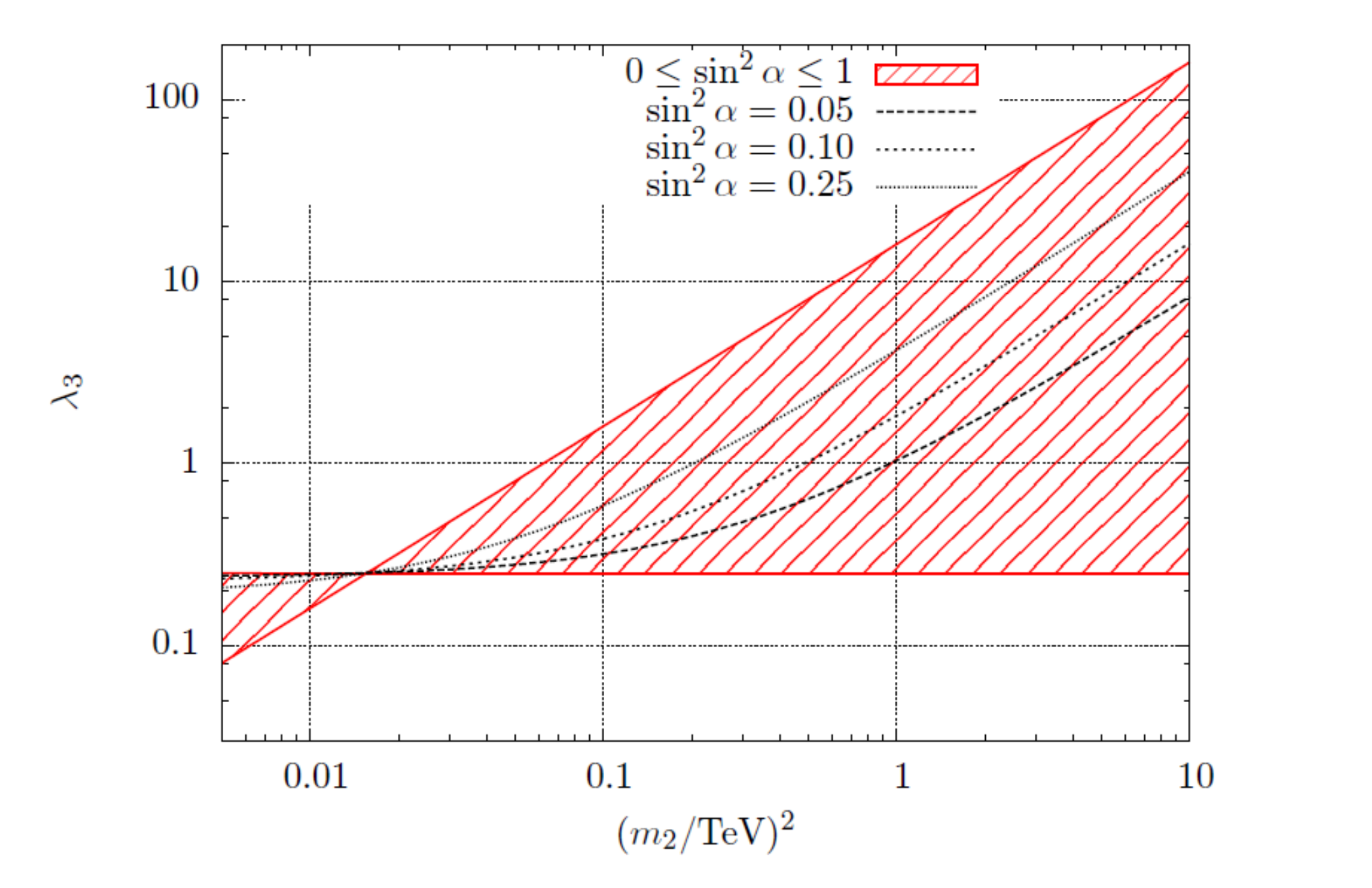}
\caption{Allowed region of parameters for the model (\ref{Lag}). The red hatched region gives the physical values (the border are not physical, but are included with the additional $\lambda_4$ coupling (see text for details)). Even for modest $\lambda_3$  most of the ($\alpha$,$m_2$) parameter space remains accessible.  \label{fig:limits}}
\end{figure}

Now let us come back to the decoupling issue. The alternative notation for equation (\ref{sinSq}):
\beq
\lambda_3 = \sin^2\alpha\left(\frac{m_2}{v}\right)^2+\cos^2\alpha\left(\frac{m_1}{v}\right)^2.\label{mixingM2}
\eeq
shows that, once $m_1$ is fixed, each mixing $\alpha$ is associated with a straight line in the plane $\left((m_2/v)^2,\lambda_3\right)$. Fig.\ref{fig:limits} shows the two extreme lines (for $m_1/v=1/2$): $\sin^2\alpha=0$ is the horizontal red line, while $\sin^2\alpha=1$ is the oblique red line. The whole set of intermediate values for the mixing corresponds to lines that lie in between (red hatched region). They all cross when $m_2=m_1$. It can be shown that the two extreme lines must be excluded from the set of physical parameters. One can understand it easily: since in this simple model the mixing is proportional to one of the diagonal elements of the mass matrix, then, if there is no mixing (extreme lines cases), one cannot find a solution for a non zero $m_2$.

The only dimensionless parameters that are relevant in the discussion are $\lambda_1$ and $\lambda_2$. Stability forces them to be positive, and then the perturbative regime is insured while $\lambda_3$ remains \textit{small}. The meaning of \textit{small} is quite subjective here (as it is difficult to judge the convergence of the series), so we simply provide a plot (Fig.\ref{fig:limits}) linking $\lambda_3$, $\alpha$ and $m_2$. Even for  very conservative values of  $\lambda_3$, a large range of parameters ($\alpha$,$m_2$) is open.

Just for completeness, the extreme values of $\sin^2\alpha=0$ and $\sin^2\alpha=1$, excluded in the basic Lagrangian, can be recovered by adding an extra term:
\beq
\Delta V=\frac{\lambda_4}{2}\left(\chi^2-\left(\frac{f_1^2}{2 f_2}\right)^2\right)^2,\label{varPot}
\eeq
with a new dimensionless parameter $\lambda_4$. This term is chosen such that the minimum of the potential is unchanged.\\
%The mixing is no longer proportional to a diagonal element of the mass matrix, and the ill-defined situation described above is avoided. It is worth mentioning that, although the expressions for the mass eigenvalues and the mixing are modified, the important formula (\ref{mixingM2}) is still valid.

In  the large $m_2$ limit,  decoupling is achieved as expected, but in a relatively slow manner (e.g. with $\lambda_3 \lesssim 2$, a mixing of 30$\%$ it still allowed for $m_2\lesssim 1$ TeV). Indeed a much faster decoupling is usually found in comparison, for instance in the \textit{Two Doublets Model (2DM)}. The more general potential for such a model can be parametrized as:
\begin{align}
V= \ &m_{11}^2\Phi_1^{\dagger}\Phi_1+m_{22}^2 \Phi_2^{\dagger}\Phi_2
-\left[m_{12}^2\Phi_1^{\dagger}\Phi_2+\text{h.c.}\right] \nonumber \\
&+\frac{\lambda_1}{2}(\Phi_1^{\dagger}\Phi_1)^2+\frac{\lambda_2}{2}(\Phi_2^{\dagger}\Phi_2)^2
+\lambda_3(\Phi_1^{\dagger}\Phi_1)(\Phi_2^{\dagger}\Phi_2)
+\lambda_4(\Phi_1^{\dagger}\Phi_2)(\Phi_2^{\dagger}\Phi_1) \\
&+\left\lbrace
\frac{\lambda_5}{2}(\Phi_1^{\dagger}\Phi_2)^2
+\left[\lambda_6(\Phi_1^{\dagger}\Phi_1)+\lambda_7(\Phi_2^{\dagger}\Phi_2)\right]\Phi_1^{\dagger}\Phi_2
+\text{h.c.}\right\rbrace. \nonumber
\end{align}
In general $m_{12}^2$, $\lambda_5$, $\lambda_6$ and $\lambda_7$ can be complex, with possible CP-violating effects. We will ignore this possibility here, as it is unrelated to our problem. If the mass matrix possesses at least one negative eigenvalue, the scalar fields will develop a VEV. Imposing CP invariance and $U(1)_{EM}$ gauge symmetry, it can be written $\langle\Phi_i\rangle=(0 \ v_i/\sqrt{2})^T$, with $v_i \in \mathbb{R}$ and $v_1^2+v_2^2=v^2$, linked to the $W$ mass. Then, it is always possible to choose a basis where\footnote{When couplings to matter are introduced, the choice of the basis is no longer arbitrary. However, this choice permits to draw more easily a parallel with the previous model and we expect our conclusions to remain valid in any basis. This will be supported by a comparison with a more general result from \cite{Haber:1995}.} $v_1=0$ and $v_2=v$.

It is then easy to derive the following relations for $m_1=$125 GeV, $m_2$ and $\alpha$, the masses and mixing of the two neutral CP-even scalars (and also $m_A$ and $m_{H^{\pm}}$ the masses of the neutral CP-odd and charged scalars that are not eaten by gauge bosons):
\begin{eqnarray}
m_2^2 \cos^2\alpha + m_1^2 \sin^2\alpha &=& m_{11}^2+\frac{1}{2}\lambda_{345}v^2\
\left(=m_A^2+\lambda_5 v^2=m_{H^{\pm}}^2+\frac{1}{2}\lambda_{45}v^2\right)  \\
m_2^2 \sin^2\alpha + m_1^2 \cos^2\alpha &=& \lambda_2 v^2 \label{sndCond} \\
(m_2^2-m_1^2)\sin\alpha\cos\alpha &=&\lambda_7 v^2,\label{thdCond}
\end{eqnarray}
where $\lambda_{ij...k}=\lambda_{i}+\lambda_j+...+\lambda_k$.

As in the ESM, there are two dimensionfull parameters $v$ and $m_{11}$ that come into the game. While the first one is fixed by the $W$ mass, the other one can be used to push all the masses (expect $m_1$) to a huge value (they all become almost equal, because the $\lambda_i$ are bounded by perturbative arguments). The relation (\ref{sndCond}) is the equivalent of (\ref{mixingM2}) in the ESM, and gives essentially the same constraint on the mixing: when $m_2$ becomes too large, a small $\sin \alpha$ must compensate.

\begin{figure}[h]
\centering
\includegraphics[scale=0.55] {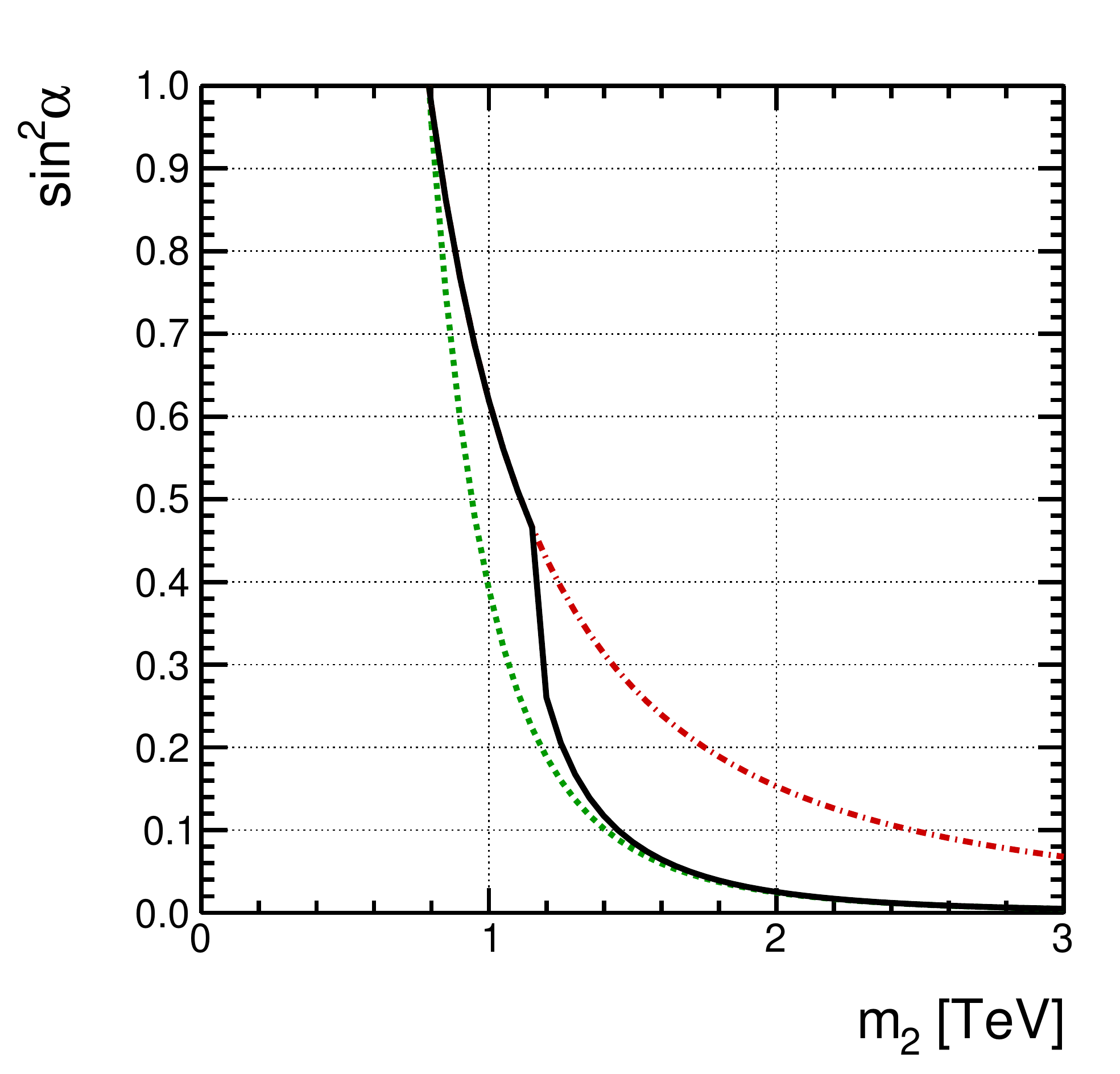}
\caption{Constraints on the mixing $\sin^2\alpha$ as a function of $m_2$ in the present ESM case (upper curve, in red) for $\lambda_3 < 10$ and in the 2DM models (upper limit in thin black, approximation (\ref{limHaber}) in green-dashed) for $\lambda_2, \lambda_7 < 10$. \label{fig:limits2HDM}}
\end{figure}

The difference in  the 2DM is the presence of the new relation (\ref{thdCond}). In the ESM, the off-diagonal term of the mass matrix contained a product of the two dimensionfull paramaters $v$ and $f_2$ rather than only $v^2$ like here. Then, when $f_2$ is increased to raise $m_2$, this term "follows" and we find no new constraint. On the contrary in the 2DM the condition (\ref{thdCond}) introduces a new constraint at higher $m_2$ as shown in Fig.\ref{fig:limits2HDM}.

In the 2DM model, this result was found in the same context,
but in the more general case where $v_1\neq 0$ \cite{Haber:1995}. The author showed that in the decoupling limit ($m_2\approx m_A\approx m_{H^{\pm}}\gg m_1$):
\beq
\cos^2(\beta-\alpha)\simeq \frac{m_L^2(m_T^2-m_L^2)-m_D^4}{m_A^4}, \label{genLimit}
\eeq
where $\tan \beta =v_2/v_1$, and $m_L$, $m_T$ and $m_D$ are some functions of the CP-even mass matrix elements (see eqns (3.5) in \cite{Haber:1995}). In our limit $\beta \rightarrow \pi/2$ and (\ref{genLimit}) turns into:
\beq
\sin^2 \alpha = \left(\frac{\lambda_7}{16 (m_2/\text{TeV})^2}\right)^2. \label{limHaber}
\eeq
We see on Fig.\ref{fig:limits2HDM} that this line gives the same constraint as (\ref{thdCond}), in the regime $m_2\gg m_1$.

\section{Combined constraints and strategy}

In this section, we return to the Extra Singlet Model as a benchmark for comparing "direct searches" to "precision measurements". While it would be conceivable to present a (presumably more constraining) fit combining the two sets of data, we rather advocate to present them separately, (i) to compare the impact of both approaches (ii) to allow for an  easy extension to more complex models (for instance, with more than one singlet).

\begin{figure}[h]
\centering
\includegraphics[scale=0.55] {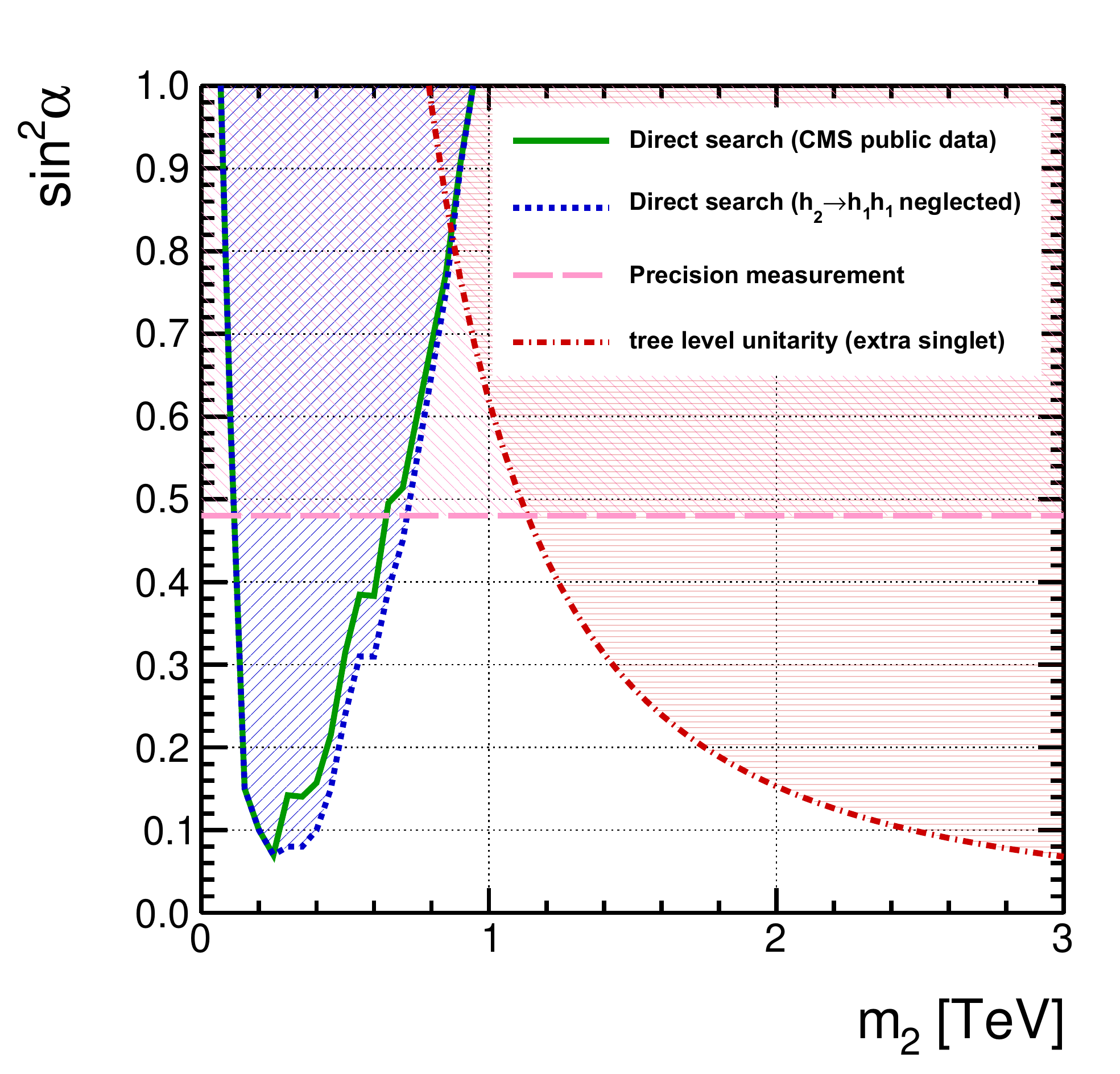}
\caption{Comparison of the constraints on the mixing $\sin^2\alpha$ as a function of $m_2$ in the ESM case. The graph shows the constraints from "perturbative unitarity" (in red, dash-dotted line) as discussed above, combined with "precision" constraints (horizontal  pink dashed line) stemming from the 125 GeV peak and "direct search" constraints (the solid  green curve gives the current limit, while the blue (dotted) curve neglects the decay $H_2\rightarrow 2H_1$; experimental constraints are inferred at  "$2\sigma$". \label{fig:param_space}}
\end{figure}

In Fig.\ref{fig:param_space} we show how these constraints restrict the parameter space (keep in mind that the "perturbative unitarity" is simply a projection of the point at which the perturbative approach to the theory is likely to fail).

\section{Conclusions}

Extensions of the Standard Model (notably through extra scalars) can be studied either by searching for a deviation in the production rate of the 125 GeV Standard Model Scalar (Brout-Englert-Higgs scalar), or by direct search for additional states. Taking the Extra Singlet Model as an example, we see that direct detection is currently much more sensitive\emph{in most of the accessible mass range }(125-600 GeV).

Taking perturbative unitarity as a guide, a full coverage of the parameter space allowed in the future by a few percent departure in production rate would require to extend the detection sensitivity to the TeV range.

The model considered here offers simplicity as a benchmark (the extra state has the properties expected from a Standard Model Scalar of the same mass, so existing analysis can be directly used). The fact that the extra component is a singlet (without its own direct interactions to known matter) makes direct detection more difficult, while the maximal mixing allowed by perturbative unitarity is rather large even for a heavy extra scalar (at least compared to the 2 Doublet Models).

\section{Acknowledgements}

This work was supported in part by IISN (Belgium) and Belgian Science Policy office (IAP VII/37); S.M., C.C. and B.C.  are respectively FNRS fellows, and "Ma\^iître de Recherches". We thank in particular P. VanLaer and J. van der Bij for constructive discussions and (for J.vdB.) for pointing out the importance of the $2H_1$ channel.

\end{document}